\documentclass[aps,pra,reprint,groupedaddress]{revtex4-1}
\newcommand{\gs}{$X^2\Sigma\,\,$}
\newcommand{\gsshort}{$X^2\Sigma$}
\newcommand{\exs}{$A^2\Pi_{1/2}\,\,$}
\newcommand{\exsshort}{$A^2\Pi_{1/2}$}
\newcommand{\deltastate}{$A'\Delta\,\,$}

\newcommand{\baf}[3]{${}^{#1#2#3}\textrm{Ba}^{19}\textrm{F}$}

\usepackage{todonotes}
\usepackage{color}
\usepackage{amsmath}

\begin{document}

\title{Buffer-gas cooling, high-resolution spectroscopy and optical cycling of barium monofluoride molecules}

\author{Ralf Albrecht}
\affiliation{5. Physikalisches Institut and Center for Integrated Quantum Science and Technology IQST, Universit\"at Stuttgart}

\author{Michael Scharwaechter}
\affiliation{5. Physikalisches Institut and Center for Integrated Quantum Science and Technology IQST, Universit\"at Stuttgart}

\author{Tobias Sixt}
\affiliation{5. Physikalisches Institut and Center for Integrated Quantum Science and Technology IQST, Universit\"at Stuttgart}

\author{Lucas Hofer}
\affiliation{5. Physikalisches Institut and Center for Integrated Quantum Science and Technology IQST, Universit\"at Stuttgart}

\author{Tim Langen}
\email{t.langen@physik.uni-stuttgart.de}
\affiliation{5. Physikalisches Institut and Center for Integrated Quantum Science and Technology IQST, Universit\"at Stuttgart}

\date{\today}

\begin{abstract}
We demonstrate buffer-gas cooling, high-resolution spectroscopy and cycling fluorescence of cold barium monofluoride (BaF) molecules. Our source produces an intense and internally cold molecular beam containing the different BaF isotopologues with a mean forward velocity of $190\,$m/s. For a well-collimated beam of \baf138 we observe a flux of more than $10^{10}$ molecules $\mathrm{sr}^{-1}$ $\mathrm{pulse}^{-1}$ in the \gsshort, $N=1$ state in our downstream detection region. Studying the absorption line strength of the intermediate \deltastate state we infer a lifetime of $\tau_\Delta=790\pm346\,$ns, significantly longer than previously estimated. Finally, highly-diagonal Franck-Condon factors and magnetic remixing of dark states allow us to realize a quasi-cycling transition in \baf138 that is suitable for future laser cooling of this heavy diatomic molecule.
\end{abstract}

\maketitle

\section{Introduction}
Molecules feature a plethora of electronic, vibrational and rotational states, together with complex interactions~\cite{Carr2009,Bohn2017}. On the one hand, this complexity is precisely what makes them so interesting. However, it simultaneously makes them also extraordinarily challenging to cool. To achieve molecular slowing and cooling, many techniques have been developed over the last decades, starting from molecular sources based on supersonic expansion or buffer-gas cooling \cite{Kantrowitz1951,Hutzler2012}, to techniques like Stark and Zeeman deceleration~\cite{Meerakker2012}, cryofuges~\cite{Wu2017}, electro-optical Sisyphus cooling~\cite{Zeppenfeld2012}, merged molecular beams~\cite{Henson2012} or electromagnetic traps~\cite{Meerakker2012,Stuhl2012,Reens2017}. Moreover, pairs of alkali atoms in an ultracold gas have been associated into diatomic molecules and coherently transferred into the rovibrational ground state to form quantum degenerate gases~\cite{Bohn2017,DeMarco2019}. However, versatile techniques that can bring chemically diverse molecular species to ultracold temperatures are scarce. In particular, many of the powerful techniques that are used to cool a large number of atomic species to ultracold temperatures have long been considered inapplicable to molecules. 

In a transformational development, laser cooling has recently been proposed and demonstrated for both diatomic and small polyatomic molecules~\cite{Rosa2004,Isaev2016,Shuman2010,Kozyryev2017}. In a series of groundbreaking experiments, trapped samples of SrF~\cite{Barry2014}, CaF~\cite{Truppe2017,Anderegg2017} and YO~\cite{Collopy2018} molecules were produced at low microkelvin temperatures. SrF and CaF were subsequently transferred into conservative potentials ~\cite{McCarron2018,Anderegg2018}, which allowed for the precise manipulation of their quantum states~\cite{Williams2018} and the demonstration of single molecule detection in optical tweezer arrays~\cite{Anderegg2019}.

\begin{figure}[tb]
\includegraphics[width=0.47\textwidth]{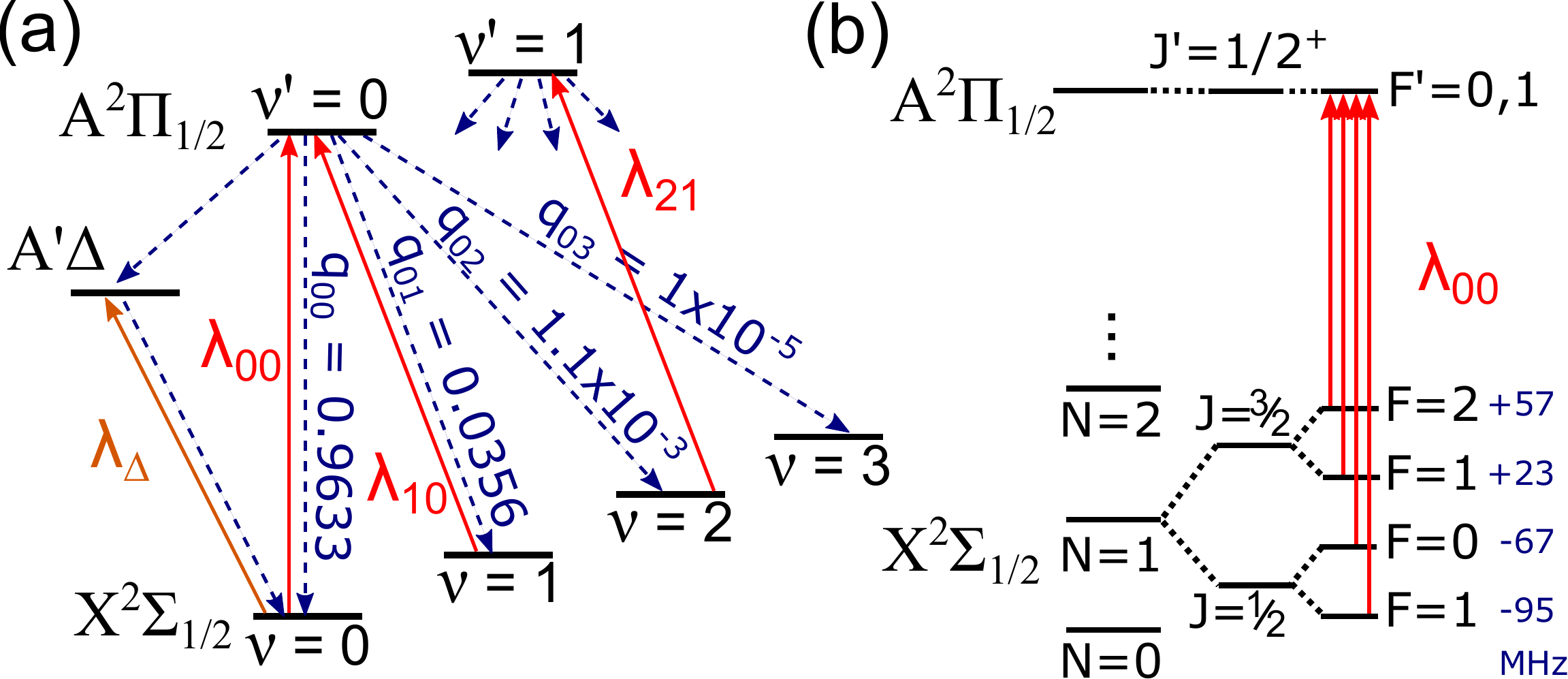} 
\caption{Energy levels and laser cooling scheme for ${}^{138}\textrm{Ba}^{19}\textrm{F}$. (a) Cooling and repumping transitions ($\lambda_{00}=859.840\,\textrm{nm}$, $\lambda_{10}=895.710\,\textrm{nm}$ and $\lambda_{21}=897.973\,\textrm{nm}$) are indicated by the red solid lines. The linewidth of the excited $A^2\Pi$ state is $\Gamma_\Pi=2\pi\times 2.84\,$MHz~\cite{Berg1998}. Decays are indicated by dashed lines, with $q_{ij}$ denoting the respective Franck-Condon factors. The branching from both the $\nu'=0$ and the $\nu'=1$ states is highly diagonal, with the precise $q_{ij}$ values for the latter given in the appendix. Overall, each molecule can scatter almost $10^{5}$ photons using only three lasers before vibrational branching into states with $\nu\geq3$ occurs. However, additional branching from the \exs into the \deltastate state can also limit optical cycling. The corresponding branching ratio has been estimated between $10^{-4}$~\cite{Hao2019} and $10^{-9}$~\cite{Kang2016}, with the precise value unknown. The linewidth of the narrow-line transition from the ground state to the $A^\prime\Delta$ state ($\lambda_{\Delta}=933.111\,\textrm{nm}$, orange solid line) is estimated in this work to be $\Gamma_\Delta = 2\pi\times (171\pm 82)\,$kHz. (b) As in CaF, SrF and YO, rotational branching can be suppressed by driving a transition from the negative parity \gsshort, $N=1$ state to the positive parity $A^2\Pi_{1/2}$, $J'^{P'}=1/2^+$ state. For optical cycling, all four ground-state hyperfine levels need to be addressed simultaneously using laser sidebands. The corresponding energy shifts are roughly equidistant and lie in the MHz range. The hyperfine structure in the excited state is not resolved.\vspace{-5pt}}
\label{fig:levelscheme}
\end{figure}

With SrF, CaF and YO being relatively light, there is a strong effort underway to extend laser cooling and trapping to heavier and more complex species like e.g. YbF~\cite{Lim2018}, YbOH~\cite{Kozyryev2017b}, TlF~\cite{Norrgard2017}, BaH~\cite{Iwata2017} and, in particular, BaF~\cite{Chen2017,Cournol2017}. The small rotational level spacing of BaF makes it easily polarizable and, together with its large mass, very well suited for precision tests of fundamental symmetries. The latter include searches for a permanent electric dipole moment of the electron~\cite{Aggarwal2018,Vutha2018,ACME2018} and parity-violating nuclear anapole moments~\cite{Altuntas2018}. For quantum simulation applications, BaF exhibits abundant bosonic and fermionic isotopologues and is doubly dipolar~\cite{Micheli2006}, with electric and magnetic dipole moments of $d=3.17\,$D and $\mu \sim 1\,\mu_B$, respectively~\cite{Ernst1986}. In addition, BaF-BaF interactions have been predicted to be predominantly elastic at low temperatures in appropriate electric fields~\cite{Gonzalez-Martinez2017}. Lastly, the dissociation threshold in BaF lies higher than the ionization threshold, leading to long-lived molecular Rydberg states~\cite{Jakubek1994,Zhou2014}. 

Here, we present and characterize our setup for buffer gas cooling of BaF, perform high-resolution spectroscopy of the relevant laser cooling transitions and study the lifetime of the $A^\prime\Delta$ intermediate excited state. Finally, in a first step towards laser cooling, we demonstrate cycling fluorescence of the resulting molecular beam using magnetic remixing of BaF's dark states. An alternative method for remixing dark ground state sublevels in BaF is polarization modulation~\cite{Chen2017}.

\section{Level structure and laser cooling scheme}

The level structure of BaF has been studied extensively spectroscopically~\cite{Effantin1990,Steimle2011} and exhibits several features that are favorable for laser cooling~\cite{Hao2019,Kang2016,Chen2016}. In this work we will focus mainly on the most abundant isotopologue \baf138, for which we summarize the laser cooling scheme in Fig.~\ref{fig:levelscheme}. All relevant transitions are located in the near-infrared spectral region, where ample laser power can easily be generated, e.g. for efficient bichromatic laser slowing and cooling~\cite{Kozyryev2018,Galica2018} or the generation of large optical forces based on mode-locked lasers~\cite{Long2019}. Favorable Franck-Condon factors between the ground \gs and excited \exs state enable the scattering of almost $10^5$ photons with only two repumping lasers, before a molecule decays to levels with higher vibrational quantum numbers $\nu \geq 3$. As in CaF, SrF and YO, the \gsshort, $N=1$ $\rightarrow$ \exsshort, $J'^{P'}=1/2^+$ transition is closed rotationally due to parity and rotational selection rules~\cite{Stuhl2008}. Here, $N$, $J$ and $P$ are rotational quantum number, total angular momentum quantum number and parity, respectively, and primes denote excited states. In addition, the nuclear spin $I({}^{19}\textrm{F})=1/2$ of the ${}^{19}\textrm{F}$ nucleus leads to a simple hyperfine structure with four hyperfine levels $F=2,1,0,1$ in the $N=1$ ground state. As the hyperfine levels $F'=0,1$ in the excited state are not resolved~\cite{Steimle2011,Chen2016}, only four transitions thus have to be addressed per vibrational level to realize optical cycling, which can conveniently be achieved by adding sidebands to the cooling and repumping lasers. The molecular structure will be discussed in further detail below and the relevant energies and branching ratios of the individual sublevels are summarized in the appendix. Notably, the $A^{2}\Pi$ excited state features a larger g-factor compared to SrF, CaF and YO, which is expected to lead to larger restoring forces in magneto-optical traps and could facilitate blue detuned magneto-optical traps~\cite{Chen2016,Tarbutt2015,Devlin2016}. 

However, laser cooling of BaF also faces several challenges. In particular, given a recoil velocity of $v_{rec}=2.94\,$mm/s on the order of $6.5 \times 10^5$ photons need to be scattered to bring a molecule from a $200\,$m/s slow beam to rest. As shown in the appendix, Franck-Condon factors in the second electronically excited state $B^2\Sigma$, which has previously been employed for slowing of CaF with improved efficiency~\citep{Anderegg2017}, are not sufficiently diagonal for optical cycling. Leakage from the \exs state into an intermediate \deltastate state may also limit conventional optical cycling, as this two-photon decay path will invert the parity of the occupied ground state. However, this effect may be less detrimental than in the YO molecule~\cite{Hao2019,Kang2016,Yeo2015}. Furthermore, the existence of the \deltastate state could also be advantageous, as a narrow-line transition is expected to connect this state and the \gs ground state. This transition may allow for the implementation of efficient stimulated slowing and cooling techniques and for lower Doppler temperatures in conventional laser cooling based on spontaneous forces~\cite{Collopy2015,Norcia2018}. 
 
\section{Experimental setup}
\begin{figure}[tb]
\includegraphics[width=0.45\textwidth]{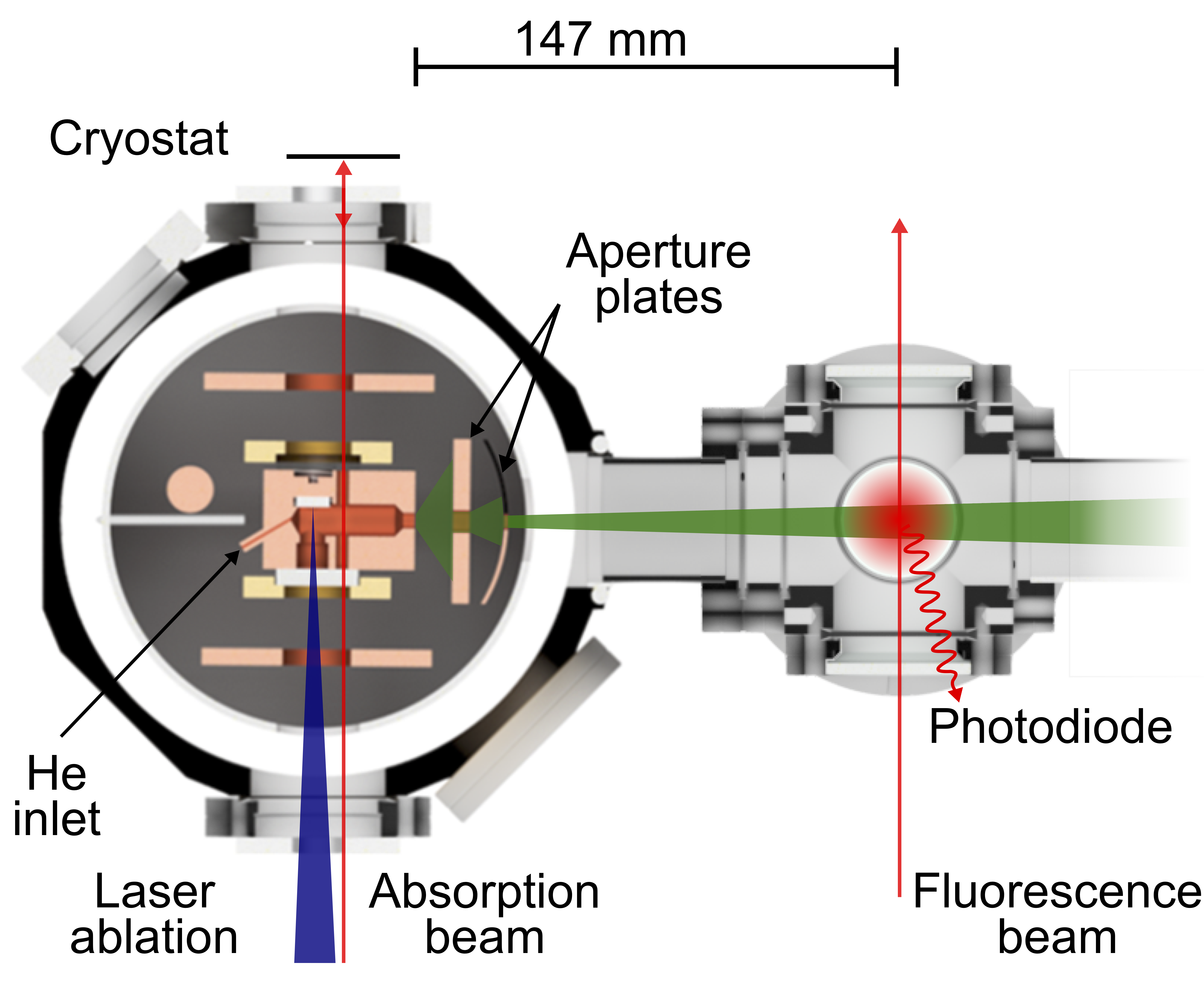} 
\caption{Experimental setup. Molecules are created by laser ablation inside a buffer gas cell, which is attached to the $4\,$K stage of a cryostat. The outcoupled slow and cold molecular beam is collimated using two cryogenic copper apertures. Probing of the molecules is possible using the absorption of a retro-reflected laser beam inside the buffer gas cell or after time-of-flight via laser-induced fluorescence.}
\label{fig:setup}
\end{figure}

Our setup to produce cold BaF molecules is depicted in Fig.~\ref{fig:setup}. Molecules are created from solid $\textrm{BaF}_2$ precursor targets by laser ablation with a $43\,$mJ per $9\,$ns pulse q-switched ND:YAG laser at $1064\,$nm, which is focused down to an ablation spot size of around $100\,\mu$m and operated at a $1\,$Hz repetition rate. The precursor target is located inside a helium buffer gas cell cooled to cryogenic temperatures using a liquid helium/liquid nitrogen ($\textrm{LHe/LN}_2$) dewar~\cite{dewar} that reaches steady-state temperatures of $4-5\,$K, depending on the buffer gas flux and the additional heat deposited by the ablation laser pulses. For a given configuration, pumping on the helium bath of the dewar using a scroll pump decreases the steady state temperature of the cryostat by around $1\,$K, which notably improves vacuum conditions and, hence, the molecular beam brightness. The dewar requires approximately $2$ hours for a cool-down or warm up, respectively. The cooling of the high purity helium buffer gas to cryogenic temperatures is ensured by two bobbins thermally anchored to the $77\,$K and $4\,$K stage of the cryostat and its flux is precisely controlled using a mass flow controller. 

The buffer gas cell is based on the design reported in Ref.~\cite{Truppe2017a}. It is machined out of a single block of copper, has inner dimensions of $40\times 50 \times 44 \,\mathrm{mm}^3$ and a conical output aperture with a diameter of $5\,$mm. The buffer gas is directed into the cell through an angled $1/8$ inch inlet in the back of the cell, a design which maximizes molecular extraction by minimizing the formation of vortices or regions with low buffer gas flow~\cite{Truppe2017a}. The buffer gas density inside the cell can be calculated for the equilibrium condition, where the flow into the cell equals the flow out of the cell~\cite{Hutzler2012}. In general, we use mass flows between $0.5$ and $5\,$sccm corresponding to buffer gas densities in the range of $n_{He} \sim 10^{14}-10^{16}\,\mathrm{cm^{-3}}$.    

Windows on both sides of the ablation cell enable optical access for laser ablation and in-cell absorption spectroscopy. The laser beam for the latter is located $9\,$mm downstream from the ablation spot. Two cryogenic apertures~\cite{Wu2018} with diameters of $7\,$mm and $5\,$mm located $13\,$mm and $24\,$mm after the cell collimate the outcoupled molecular beam to a transversal temperature of approximately $100\,$mK. After the apertures, the beam enters a room-temperature vacuum chamber. A $300\,$L/s and two $80\,$L/s turbo pumps, as well as charcoal-covered copper plates inside the cryostat, are used to remove excess helium and keep this room-temperature part of the setup at a pressure of around $10^{-6}\,$mbar. The total surface area of the charcoal is around $100\,\mathrm{cm}^2$, which allows for stable operation over runtimes exceeding $2\,$h, only limited by the refilling of the helium dewar. 

After a distance of $147\,$mm the molecular beam intersects with a second laser beam and a high-amplification photodiode is used to record the resulting fluorescence. In this region, two pairs of perpendicularly mounted coils provide magnetic fields with variable orientation and magnitude. For spectroscopy, the corresponding laser light is derived from a tunable cw Ti:Sa laser, which has a linewidth of $70\,$kHz and can be scanned modehop-free over about $20\,$GHz in the range of $700$ to $950\,$nm. Additionally, ample power is available from diode lasers at the required wavelengths for the main laser cooling transitions at $\lambda_{00}$,   
$\lambda_{10}$ and $\lambda_{21}$. 

\begin{figure}[tb]
\includegraphics[width=0.45\textwidth]{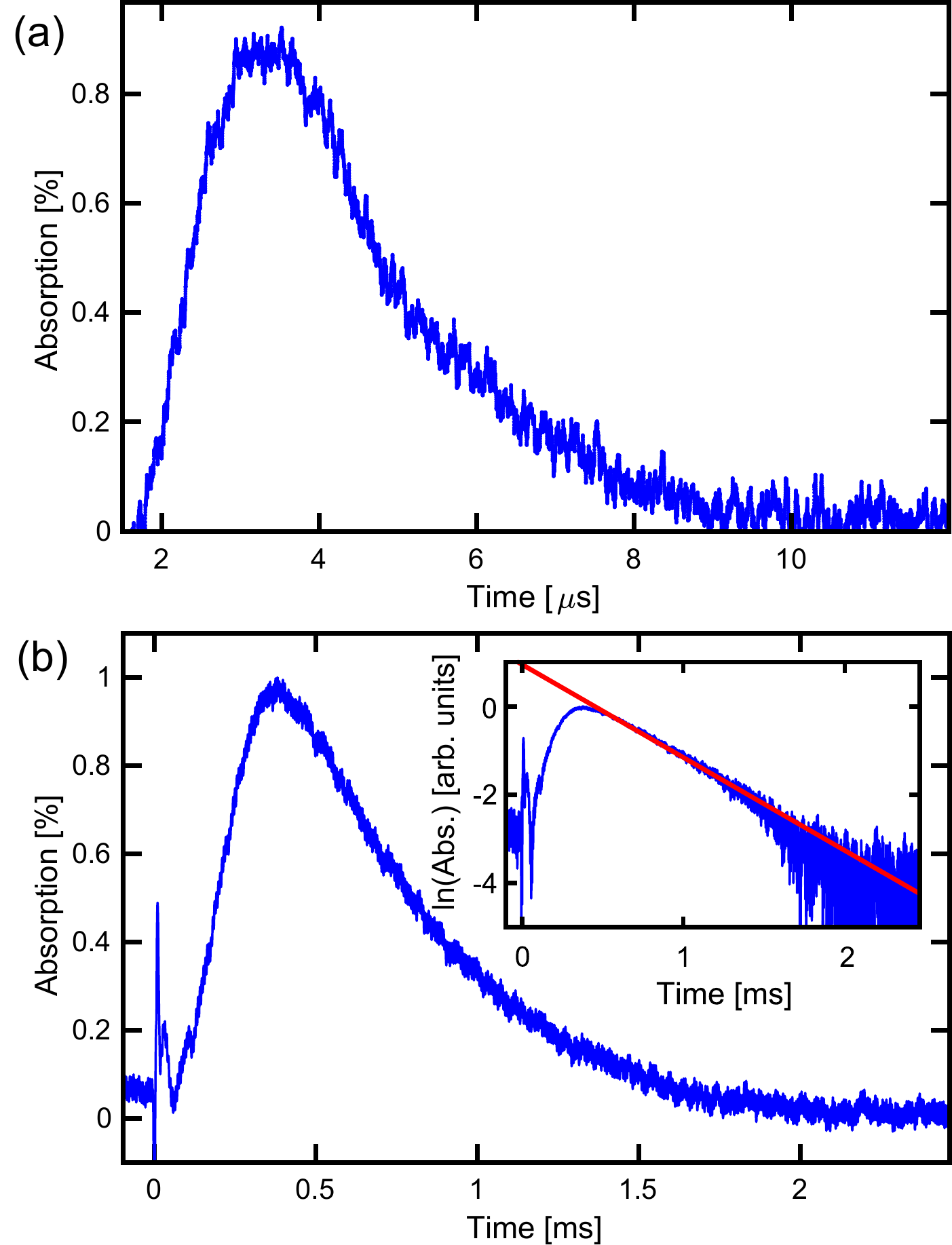} 
\caption{(a) Absorption trace of the ablation plume expanding into vacuum. The trace shown is an average over $50$ ablation pulses. (b) Molecular absorption during buffer gas cooling. The absorption initially increases as the molecules reach the probe beam, and then decays exponentially, as expected for the diffusive motion of the molecules as they collide and thermalize with the helium atoms. Note that due to the slow diffusion the timescale is now three orders of magnitude longer than in (a). Inset: Absorption in logarithmic scale. A fit (solid red line) allows us to extract a decay time of $\tau_d=0.47\,$ms, corresponding to a BaF-He collisional cross section of $\sigma_{BaF,He}=2.7\cdot10^{-14}\,\mathrm{cm}^2$.}
\label{fig:absorption}
\end{figure}

\section{Absorption spectroscopy}
In a first step, we investigate the molecules by absorption inside the buffer gas cell. 
\subsection{Ablation into vacuum}
Many diatomic radicals, including BaF, are highly reactive and thus do not exist in gaseous form under normal conditions. However, they can be created in significant numbers by laser ablation of solids containing the molecule of interest. The gas dynamics of the resulting ablation plume can be described using a hydrodynamic model~\cite{Kools1992,Tarallo2016}. This model is based on the assumption that the ablated gas cloud rapidly evaporates perpendicular to the ablation surface. After a short time the plume equilibrates through internal collisions and finally expands adiabatically into the vacuum. 

Using a laser tuned to the $X-A\,(\nu=0,\nu'=0)$ transition we can study the dynamics of the molecular cloud following ablation from the $\mathrm{BaF}_2$ target. A typical absorption trace for a laser power of $\sim 300\,\mu$W, well below saturation, is shown in Fig. \ref{fig:absorption}a. The expanding molecular cloud passes the probe laser beam on a microsecond timescale. From this we estimate typical ablation temperatures of several thousand degrees Kelvin, in agreement with previous observations at the given ablation laser fluence of around $500\,\mathrm{J}/\mathrm{cm}^2$~\cite{Tarallo2016}.

We find the the highest molecular yield for amorphous precursor targets produced following the procedure outlined in Ref. \cite{Zhou2014}. Typically, a mixture of $95\%$ $\textrm{BaF}_2$ and $5\%$ $\textrm{CaF}_2$ is ground to a fine powder, pressed to a $1\,$cm diameter, $3\,$mm thick pellet, which is compressed using a hydraulic press and subsequently sintered in a vacuum furnace. Higher amounts of $\textrm{BaF}_2$ slightly increase the molecular yield, but reduce the cohesiveness of the pellet. The final density of the pellets is around $\sim 85\%$ of the density of pure crystalline $\textrm{BaF}_2$. For comparison, we have also investigated the use of flat, macroscopic pieces of pure, crystalline $\textrm{BaF}_2$, as used in Ref.~\cite{Iwata2017} for the creation of BaH from $\mathrm{BaH}_2$. However, we find that these exhibit unreliable yield, short lifetimes and will often develop macroscopic defects, as previously reported also in Ref.~\cite{RahmlowPhD}. We attribute this behavior to the transparency of the crystalline $\textrm{BaF}_2$ at the ablation laser wavelength. Successful ablation thus requires inefficient multi-photon processes, leading to temperature gradients, which rapidly destroy the crystalline structure of the pieces. 

Repeating our experiment many times for the same ablation spot we characterize the lifetime of our targets and find an approximately linear decay with a decay constant of around $1500$ shots. Due to inhomogeneities in the mixture of the ground $\mathrm{BaF}_2$ and $\mathrm{CaF}_2$ powder in terms of particle size and local concentration, the lifetime and peak absorption varies from spot to spot, which is also reflected in a varying maximum absorption strength for the different measurements presented in the following. The longest lifetimes are consistently observed for ablation targets with rough surfaces, which we produce by splitting one sintered pellet in half.

\begin{figure*}[tb]
	\includegraphics[width=0.99\textwidth]{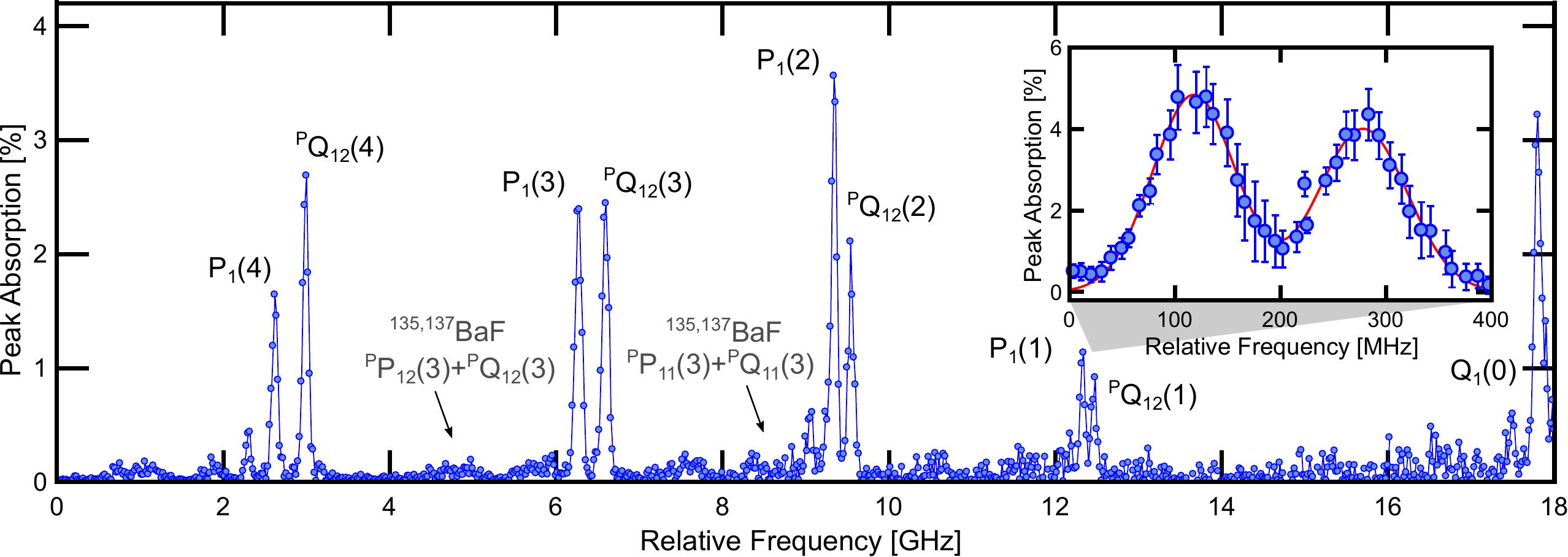}
	\caption{High-resolution spectroscopy of the \gs-$A^2\Pi_{1/2}$ band in the spectral region relevant for laser cooling. The strongest lines are due to the \baf138 isotopologue. The $P_1(1)$ and ${}^PQ_{12}(1)$ lines correspond to the main laser cooling transitions $\lambda_{00}$. Specifically, these lines result from transitions of the $N=1$, $J=3/2$ and $J=1/2$ ground states, respectively, to the $J'=1/2^+$ excited state. A scan of these transitions recorded with an increased resolution is shown in the inset. The hyperfine structure of these states is not resolved due to a residual Doppler broadening of $\sim 60\,$MHz at $5\,$K. A large number of weaker lines of the less abundant isotopologues are also visible in the spectrum. As an example, we have labeled the $G=1$ and $G=2$ components of the ${}^PP_{1G}(3)+{}^PQ_{1G}(3)$ transitions of the \baf135 and \baf137 isotopologues, which we investigate in more detail in Fig.~\ref{fig:highres_spectra}.}
	\label{fig:largespectrum}
\end{figure*}

\subsection{Buffer-gas cooling}
Due to the high temperature of the ablated molecules, only a vanishing fraction of them is found in the states involved in the laser cooling scheme discussed above. We thus introduce cold helium gas into the cell to perform buffer gas cooling~\cite{Hutzler2012}. 

With the buffer gas present, instead of expanding freely the hot molecules collide both elastically and inelastically with the cold helium atoms. After a short phase of rapid ballistic expansion, this leads to a diffusive motion towards the cell walls, during which both the translational and rotational degrees of freedom of the molecules are cooled. The timescale of the absorption signal correspondingly increases from microseconds to milliseconds due to the slowing of the expansion by the buffer gas. An example absorption signal is shown in Fig~\ref{fig:absorption}b. 

The decay time $\tau$ of the molecular absorption signal is directly related to the diffusion time $\tau_d$ by  $1/\tau = 1/\tau_{d}+1/\tau_{p}$, which allows us to extract the collisional cross section $\sigma_{He,BaF}\sim\tau_d$ for the He-BaF system. 
Here, $\tau_{p}=V/C$ is the extraction or \textit{pump out} time of the buffer gas and molecules out of the cell, which depends on the cell volume $V$ and the conductance $C$ of the cell aperture. Assuming molecular flow, this conductance is given by $C=A_{aperture}\bar{v}_{He}/4$~\cite{Steckelmacher1966}, where $\bar{v}_{He}=\sqrt{8k_B T/\pi m_{He}}$ is the mean thermal velocity of the helium buffer gas, $A_{aperture}$ is the cross section of the cell aperture and $k_B$ is the Boltzmann constant. For the example shown in Fig.~\ref{fig:absorption}b we extract a decay time of $\tau = 0.47\,$ms by fitting an exponential to the tail of the molecular absorption trace. Based on our geometry, we estimate a pump out time of $\tau_{p}=3.65\,$ms. 
Together with the buffer gas flow of $1.3$ sccm used in this measurement, this yields a diffusion time of $\tau_{d} = 0.55\,$ms and a cell parameter $\gamma_{cell}=\tau_{d}/\tau_{p}\approx 0.15$. Following Ref.~\cite{Hutzler2012}, we approximate the collisional cross section by
\begin{equation}
	\sigma_{BaF,He}=\frac{9\pi\bar{v}_{He}\tau_{d}}{16 A_{cell} n_{He}},
\end{equation}
where $A_{cell}$ is the characteristic cell cross section of the cell. Here we have assumed $n_{He}\gg n_{BaF}$ and $m_{BaF}\gg m_{He}$. With this we find a collisional cross section of $\sigma_{BaF,He}=2.7\times 10^{-14}\,\mathrm{cm}^2$. This result is in good agreement with previous results for the He-BaF system~\cite{Chen2017} and underlines the suitability of BaF for buffer gas cooling. Note that the uncertainty of this result based on the experimental uncertainties is negligible and hence not given here. However, significant systematic uncertainties are expected, as the result above relies on a simplified diffusion model, which does not fully capture the complex dynamics inside the buffer gas cell. For a future absolute measurement, the influence of the cell geometry and local helium density could be calibrated precisely using a simple reference atom like lithium, for which the cross-section with the buffer gas is well known~\cite{Skoff2011}.

\subsection{High-resolution spectroscopy}
In a next step, we perform high-resolution spectroscopy of the cold BaF molecules. For this, the frequency of the in-cell probe laser is scanned and the strength of the molecular absorption is evaluated from the absorption trace for each frequency. The frequency range is chosen to cover the lines around the cooling transition at $\lambda_{00}$ in the $X-A\,(\nu=0,\nu'=0)$ band. Over the scan range, the ablation spot is changed frequently to realize approximately constant conditions. In addition, each data point is averaged over $8$ individual absorption traces, leading to $1200$ shots per ablation spot. 

The result is shown in Fig.~\ref{fig:largespectrum} and shows lines from many of the various BaF isotopologues. As ${}^{19}\textrm{F}$ is the only stable fluorine isotope, their abundances - and therefore the corresponding linestrengths - are determined by the isotopic abundances of atomic Ba, leading e.g. to bosonic \baf138 ($71.70\%$) and \baf136 ($7.85\%$), and fermionic \baf137 ($11.23 \%$) and \baf135 ($6.59\%$). In the following, we briefly summarize the relevant details of the molecular structure to describe the most prominent of the observed lines. For the corresponding molecular constants we refer the reader to Ref.~\cite{Effantin1990}.

The \gs ground state of BaF is described by Hund's case (b), where the total angular momentum $\mathbf{J}=\mathbf{S}+\mathbf{N}$ is given by the coupling of electron spin $\mathbf{S}$ and rotational angular momentum $\mathbf{N}$~\cite{Steimle2011}. For each rotational level $N$ with parity $(-1)^N$, this leads to pairs of fine structure states with $J=N\pm 1/2$, respectively. Theses pairs are split by $\gamma(N+1/2)$, where $\gamma = 81.68\,$MHz is the spin-rotation constant. For the even isotopologues like \baf138, coupling of $\mathbf{J}$ and $\mathbf{I}({}^{19}\textrm{F})$ further leads to the total angular momentum $\mathbf{F}=\mathbf{J}+\mathbf{I}({}^{19}\textrm{F})$, and thus a splitting of each of the two fine structure levels into two hyperfine levels. The result are the four hyperfine states $F=N+1,N,N,N-1$ per rotational level, which are, however, not individually resolved in our spectroscopy due to the residual Doppler broadening at $5\,$K. 

The excited $A^2\Pi$ state is described by Hund's case (a), and splits into $A^2\Pi_{3/2}$ and $A^2\Pi_{1/2}$, which are well separated in energy by a spin-orbit coupling of $18.96\,$THz~\cite{Effantin1990}.
The total angular momentum $\mathbf{J}$ is a good quantum number and formed by coupling the total electronic angular momentum $\mathbf{\Omega}$ with the nuclear rotational angular momentum. For each $J'$, $\Lambda$-doubling results in a pair of sublevels with opposite parity $P'$. The hyperfine structure with $F'=0,1$ in this state is unresolved.

Such $\Sigma-\Pi$ transitions are typically labeled using the notation $^{\Delta N}\Delta J_{a,b}(N)$~\cite{Herzberg}. Here, $\Delta N=N'-N$ and $\Delta J=J'-J$, with the primed and unprimed quantum numbers again referring to the excited and ground state, respectively. The parameter $a=1,2$ denotes the parity sublevel of the $A^2\Pi$ state involved in the transition, the parameter $b$ is $1$ if $J = N+1/2$ and $2$ if $J = N-1/2$ in the involved \gs state. If $a = b$ or $\Delta J = \Delta N$ the repeated label is omitted. As usual $P$, $Q$ and $R$ denote transitions changing angular momentum quantum numbers by $-1$,$0$ or $1$, respectively. 

\begin{figure}[b]
\includegraphics[width=0.46\textwidth]{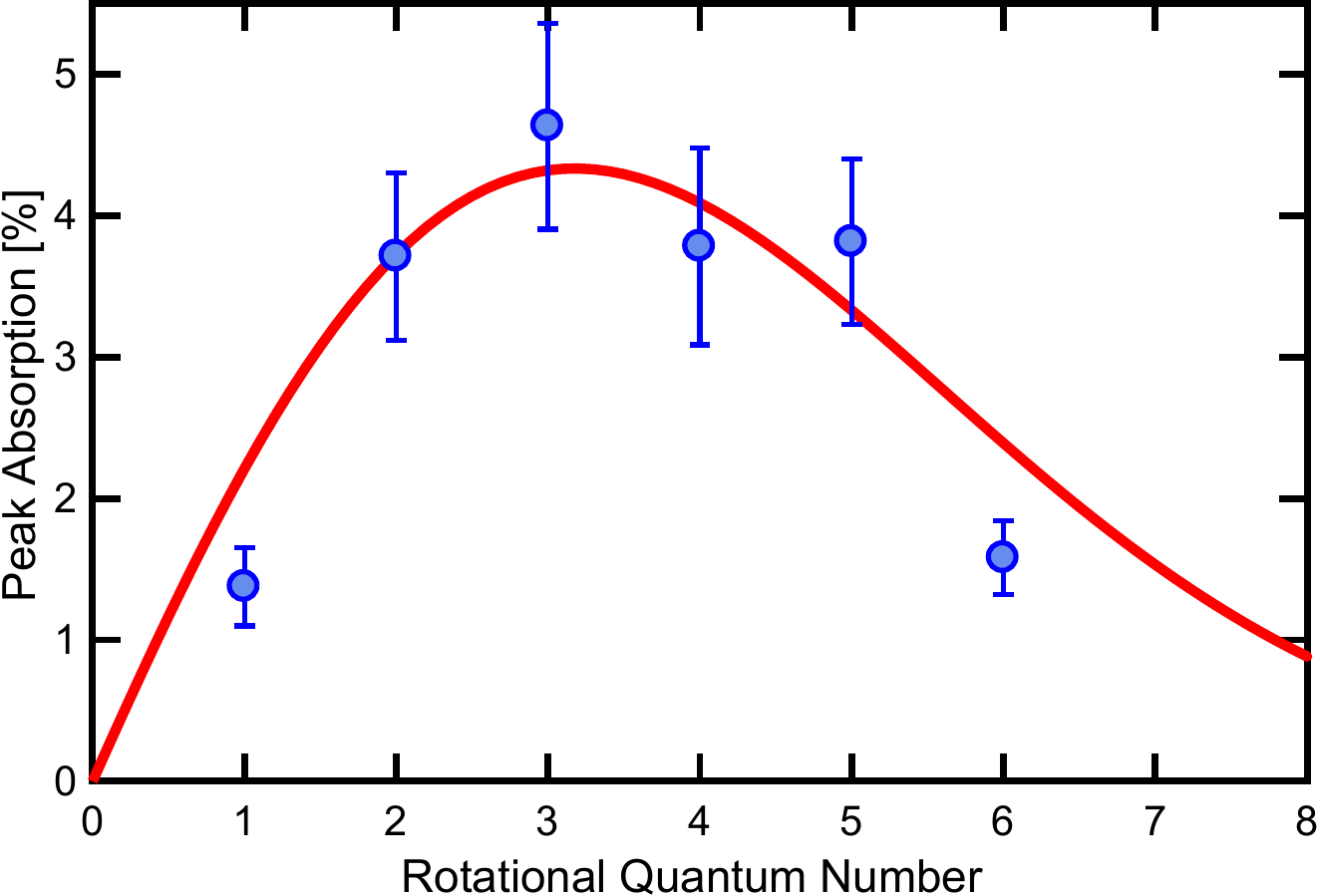}
\caption{Peak absorption of the $^{P}Q_{12}(N)$ branch versus the rotational quantum number $N$. The red line is a fitted Boltzmann distribution, with a temperature of $T_{rot}=7.2\pm 3.1\,$K.}
\label{fig:rotationalcooling}
\end{figure}

Based on this we can identify the lowest lying $P_1(N)$ and ${}^PQ_{12}(N)$ transitions of \baf138, as well as the $Q_1(0)$ transition as the dominant lines in Fig.~\ref{fig:largespectrum}. As expected, the splitting of $P_1(N)$ and ${}^PQ_{12}(N)$ increases proportional to $\gamma$ with increasing $N$.  Most importantly, the $P_1(1)$ and ${}^PQ_{12}(1)$ lines connect the \gsshort, $N=1,J=3/2$ and $J=1/2$ states with the \exsshort, $J'^{p'}=1/2^+$ state and thus correspond to the main cooling transitions of the laser cooling scheme discussed above. A more detailed measurement of these transitions is shown in the inset of Fig.~\ref{fig:largespectrum}. The absolute transition frequency agrees well with the calculated value $\lambda_{00}=859.840\,$nm within the accuracy of our frequency reference~\cite{Effantin1990,Steimle2011}.

A measurement of the rotational temperature of the molecules can be obtained from the mean absorption of the lines in the $^{P}Q_{12}(N) $ branch, which reflects the Boltzmann distribution of the populations of the various rotational levels in the ground state. To minimize systematic effects from the depletion of the ablation target, we chose a new target spot for every investigated line from $N=1$ to $N=6$. The result is shown in Fig.~\ref{fig:rotationalcooling}. A fit to the data using a Boltzmann distribution, including the degeneracy factor of $4N$ per rotational level for $^{P}Q_{12}(N)$ transitions, yields a temperature of $T_{rot}=7.2\pm 3.1\,$ K, which is compatible with the $5\,$K measured for the cryostat.

The spectra for the other two, less abundant even isotopologues \baf136 and \baf134 are much weaker, but identically structured. For the given transitions, they are shifted to lower frequencies by around $600\,$MHz and $750\,$MHz for \baf136 and \baf134, respectively~\cite{Steimle2011}. 

Next, we discuss the structure of the odd, fermionic isotopologues such as \baf137 and \baf135, which is complicated by the additional nuclear spin $I({}^{137,135}\mathrm{Ba})=3/2$ of the odd barium nuclei. The corresponding hyperfine splitting is comparable in magnitude to the rotational splitting and thus significantly convolutes the spectra. The ground state can be described by a sequential Hund's case ($b_{\beta S}$) coupling scheme, with the intermediate angular momenta $\mathbf{G}=\mathbf{S}+\mathbf{I}({}^{135,137}\mathrm{Ba})$ and $\mathbf{F}_1=\mathbf{N}+\mathbf{G}$, and the total angular momentum $\mathbf{F}=\mathbf{I}({}^{19}\mathrm{F})+\mathbf{F}_1$. As in the even isotopologues, the hyperfine structure contributions to the excited state are negligible. 

\begin{figure}[tb]
\includegraphics[width=0.44\textwidth]{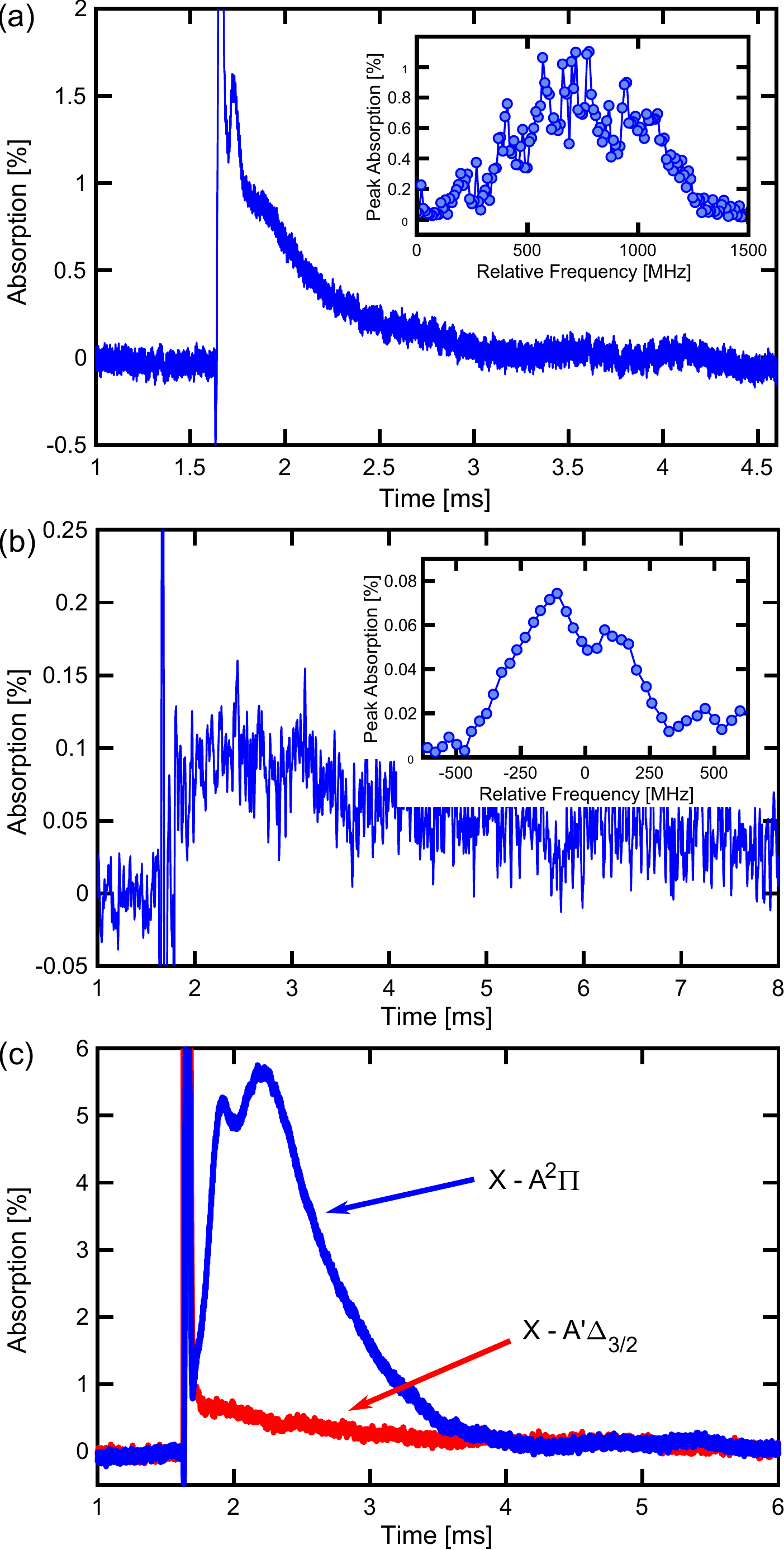}
\caption{(a) Absorption in the region of the $^{P}P_{12}(3)+{}^{P}Q_{12}(3)$ transition. Inset: High-resolution spectrum, containing many overlapping hyperfine components from \baf137 and \baf135. (b) Absorption signal at the repumping transition $X(\nu=1)\rightarrow A(\nu'=0)$ of \baf138. Insets: High-resolution spectrum, showing the fine structure splitting of the transition into $P_1(1)$ and ${}^PQ_{12}(1)$. (c) Examples of absorption signals on the X-\exs and X-\deltastate transitions, as used in the analysis of the \deltastate state lifetime. The peak around $1.6\,$ms in all panels is caused by noise from the ablation laser. It is strongest in (c) due to less efficient filtering of the $1064\,$nm ablation light, when probing the $933\,$nm $X-\Delta$ transition.}
\label{fig:highres_spectra}
\end{figure}

Besides many other transitions, which are analyzed in detail in Ref.~\cite{Steimle2011}, the $P_1(N)$ and ${}^PQ_{12}(N)$ transitions discussed for \baf138 now overlap and form a new branch. This branch is commonly denoted by $^{P}P_{1G}(N)+{}^{P}Q_{1G}(N)$. Note that the parameter $b$ in the transition labeling scheme now corresponds to the two possible values for the good intermediate quantum number $G$~\cite{Steimle2011}. Accordingly, this branch splits into two components $G=1,2$, each containing several overlapping hyperfine transitions from both \baf137 and \baf135. As indicated in Fig.~\ref{fig:largespectrum}, these two sub-branches are shifted by around $+3\,$GHz and $-1.5\,$GHz, respectively, with respect to the position of the corresponding \baf138 lines. A high-resolution spectrum of the $^{P}P_{12}(3)+{}^{P}Q_{12}(3)$ transition of \baf137 and \baf135 is shown in Fig.~\ref{fig:highres_spectra}a.

Another important ingredient for the laser cooling scheme is the $\lambda_{10}$ repumping transition, which we investigate in Fig.~\ref{fig:highres_spectra}b. As previously reported for SrF~\cite{Barry2011} and BaF~\cite{Bu2017}, we also detect a finite absorption on this line, corresponding to a small population of molecules in the $\nu=1$ vibrational state. While the peak absorption is $55\pm30$ times lower than on the $\lambda_{00}$ cooling transition, it is still significantly higher than expected from thermal equilibrium at $5\,$K~\cite{Bu2017}. Our observation is thus in agreement with the notion that vibrations are much slower to thermalize in buffer gas cooling than other degrees of freedom~\cite{Hutzler2012}. Again, we find the absolute position of this transition to be in good agreement with the existing spectroscopic data yielding $\lambda_{10}=895.710$~\cite{Effantin1990}. 

\subsection{$X-A'\Delta$ transition}
A particularly important feature for the laser cooling of BaF is the intermediate \deltastate state. The lifetime of this state is an important parameter to consider for two reasons. First, for efficient optical cycling any leakage from the \exs state to the \deltastate state should, in the ideal case, rapidly decay back to the ground state. Second, as in YO, and contrary to the above requirement, the transition from the \gs state to the \deltastate state is expected to be narrow, which could in the future be used for Doppler cooling to lower temperatures~\cite{Collopy2015} or for cooling schemes based on adiabatic stimulated forces~\cite{Norcia2018}. The forces in the latter case could significantly exceed spontaneous forces, and would thus be a powerful tool for the slowing and cooling of heavy molecules such as BaF. However, so far the lifetime of the \deltastate state has not been measured.

The decay from this state to the ground state is dipole forbidden and only arises due to mixing between the \deltastate and the various $A^2\Pi$ states. As with the $A^2\Pi$ states, the \deltastate state is described by a Hund's case (a) coupling scheme with $J$ being a good quantum number. Following the same procedures as for the previous lines, we have studied the absorption on the \gsshort, $N=2$, $J=3/2$ $\rightarrow$ $A'\Delta_{3/2}, J'^{P'}=3/2^-$ transition. The $\Lambda$-doubling of the latter state is small and not resolved in our measurement. Again, we find the absolute transition wavelength to be in good agreement with the existing molecular constants, which yield $\lambda_{\Delta}=933.111\,$nm~\cite{Barrow1988}. Comparing the relative peak absorption $A_\Delta=0. 6\pm 0.2\,$\% on this transition with the absorption $A_\Pi=5.5 \pm 1.5\,$\% on the \gsshort, $N=1$, $J=1/2$ $\rightarrow$ \exs, $J'^{P'}=1/2^+$  transition, we can estimate the lifetime of the \deltastate state from the known lifetime $\tau_{\Pi}=56\pm0.1\,$ns of the \exs state~\cite{Berg1998}. Fig.~\ref{fig:highres_spectra} shows example absorption traces from this measurement. 

The $A'\Delta_{3/2}$ state mixes predominantly with the $A^2\Pi_{3/2}$ state and therefore we assume that the transition line strengths from the \gs state to the \deltastate state are of normal $\Sigma-\Pi$ character~\cite{Simard1992}. As both transitions probed are $Q$ transitions, their H\"onl-London factors are identical~\cite{Herzberg,RahmlowPhD}. With this, we find
\begin{equation}
\tau_{\Delta}=\frac{\lambda_{\Delta}^3}{\lambda_{\Pi}^3}\frac{p(N=2)}{p(N=1)}\frac{A_\Pi}{A_\Delta}\frac{q_\Delta}{q_\Pi}\times \tau_{\Pi}.
\end{equation}
Here, $\lambda_{\Delta,\Pi}$ denotes the transition wavelengths, $q_\Delta/q_\Pi\approx0.9$ is the calculated ratio of the Franck-Condon factors, and $p(N=2)/p(N=1)\approx 1.3$ is the ratio of the Boltzmann distributed populations in the involved $N=2$ and $N=1$ levels of the \gs state at $5\,$K. With this we find a lifetime of $790\pm 346\,$ns, or alternatively a linewidth of $\Gamma_\Delta=2\pi\times (201\pm 88)\,$kHz. This lifetime is significantly longer than the $\tau_{\Delta,mix}=220\,$ns previously estimated theoretically from the mixing of the \deltastate state with other nearby states~\cite{Chen2016}. Rather, it is consistent with recent ab-initio calculations that predict a much longer lifetime of up to $5\,$ms~\cite{Hao2019}. In this context it is important to note that collisions with the buffer gas are expected to systematically decrease the lifetime of the excited states inside the buffer gas cell. We thus interpret our measurement as a lower bound for the true lifetime in free space. Based on our current estimate Doppler cooling on this transition would lead to a temperature $T_{Doppler}=\hbar \Gamma_\Delta/2k_B\approx 5 \,\mu$K~\cite{Collopy2015}, which is comparable to temperatures reached in gray molasses cooling~\cite{Truppe2017} and sufficiently cold for the direct transfer into an optical dipole trap. Here, $\hbar$ is the reduced Planck constant. Moreover, it is particularly noteworthy that our estimate is comparable to the lifetime of the narrow $626\,$nm transition in dysprosium with $\Gamma_{Dy}=2\pi\times136\,$kHz, where efficient slowing based on stimulated sawtooth-wave adiabatic passage has recently been demonstrated~\cite{Petersen2018}.

\begin{figure}[tb]
\includegraphics[width=0.44\textwidth]{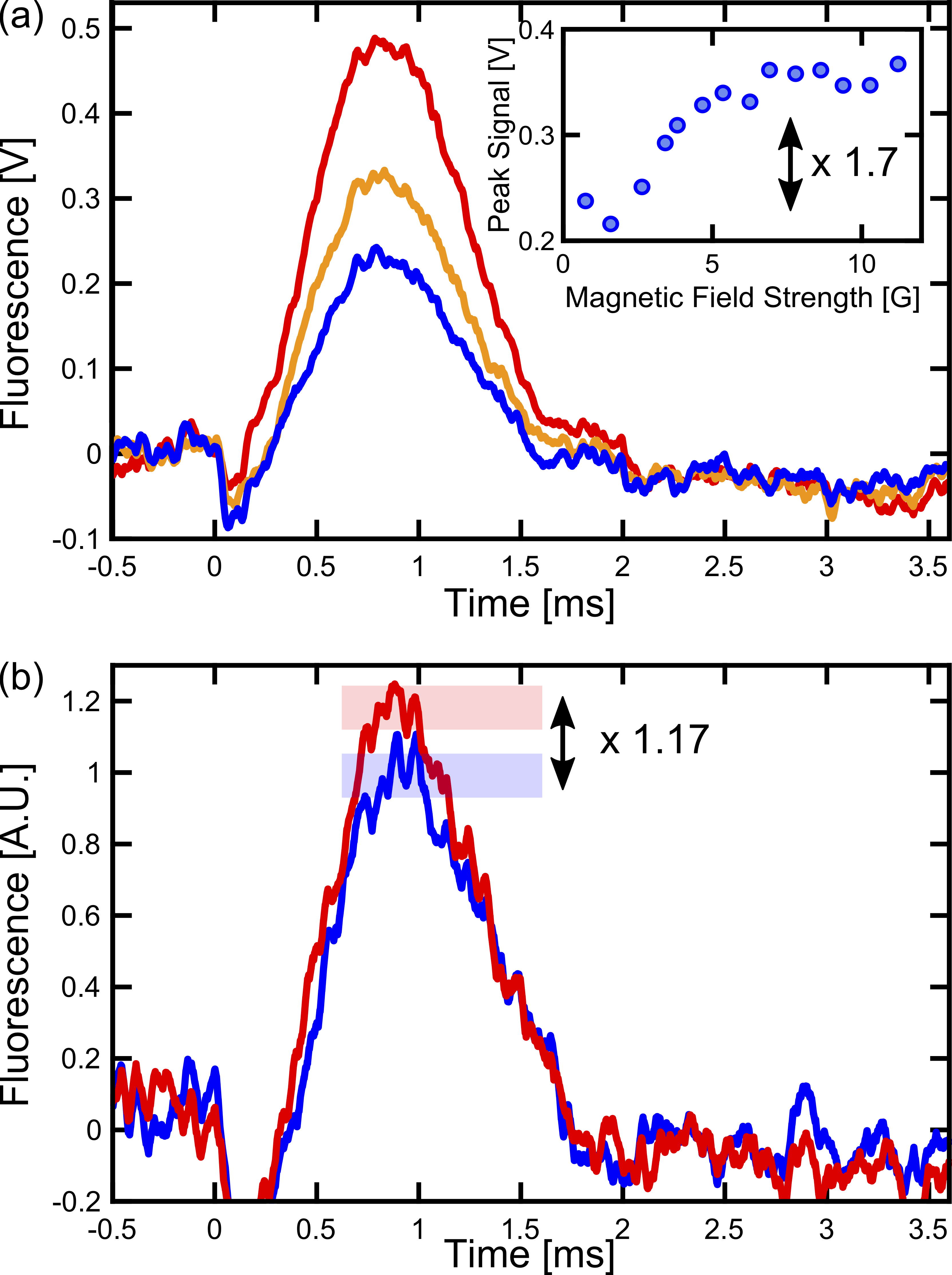}
\caption{(a) Remixing of dark states enabled by magnetic fields. Laser-induced fluorescence of the molecular beam passing a transversal laser beam resonant with all four hyperfine transitions. A magnetic field is used to remix dark states, which increases the observed amount of fluorescence. The magnetic fields are $8.64\,$G, $3.43\,$G and $0.76\,$G from top to bottom. In all traces a constant fluorescence offset from residual ablation laser light scatter has been subtracted. Inset: Peak amplitude of the fluorescence signal as a function of the applied magnetic field. The signal saturates at a field of $\sim5\,$G, where the associated Larmor frequency corresponds to around twice the excited state decay rate. The peak fluorescence increases by a factor of $1.7$ with respect to the non-remixed case at zero field. (b) The addition of a repumper leads to a further increase of the fluorescence signal by a factor of $1.17$.}
\label{fig:remix}
\end{figure}

\section{Molecular beam}
Following buffer gas cooling, the molecules exit the buffer gas cell and form an intense molecular beam. This beam is first collimated and then enters a downstream detection region where we probe it by laser induced fluorescence. Repeating the spectroscopy of the BaF transitions using laser induced fluorescence we find the laser cooling transition frequencies and rotational temperatures to be consistent with our in-cell absorption spectroscopy. In this single-frequency configuration only a single photon is scattered per molecule~\cite{Chen2017}. Tuning our laser to some of the $^{P}P_{12}(3)+{}^{P}Q_{12}(3)$ transitions that were presented in Fig.~\ref{fig:highres_spectra}, we also observe clear fluorescence signals from the odd \baf137 and \baf135 isotopologues. 

\subsection{Optical cycling}
Returning to \baf138, we then add sidebands to our laser beams using an electro-optical modulator (EOM) driven at $39.3\,$MHz to address all four hyperfine levels in the $N=1$ state simultaneously. This approach using a single EOM is facilitated by the symmetric hyperfine shifts of the ground state manifold in BaF (see Fig. \ref{fig:levelscheme}). In this configuration, transitions with $F>F'$ are possible. This leads to dark Zeeman sublevels in the ground state, which can not be addressed by our linear polarized laser beam. To remix these dark states we apply a tunable magnetic field under an angle of $\pi/4$ relative to the polarization axis of the laser beam. This scheme has been analyzed in detail in Ref.~\cite{Berkeland2002} and has successfully been used to remix dark states in a number of other molecules~\cite{Shuman2009,Lim2018,Kozyryev2018}. 

The results are shown in Fig.~\ref{fig:remix}. We observe a monotonous increase of the fluorescence signal strength with increasing magnetic field. As soon as the Larmor frequency  associated with the tilted magnetic field exceeds the linewidth of the transition the fluorescence signal starts to saturate. In principle, even higher magnetic fields could eventually lead to a decrease of fluorescence~\cite{Kloeter2008}. However, due to the small field strengths employed, the complexity of the molecular cycling transition and the broadening of the excitation laser, this effect does not play a role here.

Overall, the remixing increases the signal by a factor of approximately $1.7$. Further adding the first repumping laser at $\lambda_{10}$, including sidebands to address all hyperfine levels in $\nu=1$, increases the fluorescence by another factor of $1.17$. In our present setup this enhancement of fluorescence is limited by the interaction time between the molecules and the laser beam. The populations initially hidden in the corresponding dark states can be estimated to be $16.6\%$ for the magnetic sublevels ($2$ dark states out of $12$ ground states for linearly polarized light) and $1.4-4\%$ for the $\nu=1$ state (based on the absorption strength measurement presented in Fig.~\ref{fig:highres_spectra}b). The observed increase for both magnetic remixing and the addition of a repumper significantly exceeds these initially hidden populations, indicating successful optical cycling.

\subsection{Mean velocity and beam flux}

From the arrival time of $0.865\,$ms with respect to the ablation pulse, we can estimate the mean forward velocity of the molecular beam. Assuming an outcoupling time of $100\,\mu$s for this type of buffer gas cell~\cite{Truppe2017a}, and given the distance of $147\,$mm between the cell aperture and the detection region, we find a mean velocity of $190\,$m/s for a helium flow of $1.3$ sccm. Other velocities can be obtained by changing the buffer gas density. Given the $1\,$mm diameter of our fluorescence laser beam, this corresponds to an interaction time of around $t_{int}=5\,\mu$s between the laser and the molecules. 

A rate equation model assuming perfect magnetic remixing and taking into account all involved levels and their branching ratios captures the dynamics well~\cite{Chen2017}. In saturation, the scattering is only limited by the interaction time $t_{int}$ and the molecules scatter $\Gamma_\Pi \times N_e/(N_g+N_e)\, t_{int}= 12\pm3$ photons. Here, $N_e=4$ is the number of excited states and $N_g=24$ is the number of ground states involved in the $\lambda_{00}$ and $\lambda_{10}$ transitions. Given the quantum efficiency of around $70\,$\% of our photodiode at $860\,$nm, we deduce a number of $N_{photon}=6.75\times 10^4$ detected photons. Taking into account the numerical aperture NA=$0.2$ of the objective, the transmission through the optical elements, the fluorescence laser beam diameter and the distance from the cell, we obtain a detection efficiency of $0.34\,$\%. With this we find the number of $N=1$ molecules in our beam to be  $(4.6\pm2.7)\times10^{10}\,\mathrm{sr}^{-1}\mathrm{pulse}^{-1}$. This also allows us to estimate a number of $\sim 10^9\,\mathrm{sr}^{-1}\mathrm{pulse}^{-1}$ \baf135 or \baf137 molecules in $N=1$. Our well-collimated beam thus provides a highly intense source of internally cold molecules for a large variety of further experiments.

\section{Conclusion}
We have presented our setup for the production and spectroscopic study of cold BaF molecules. 

The flux of the less abundant \baf137 and \baf135 isotopologues in our beam is comparable to the flux of the much more abundant \baf138 used in the recent demonstration of a sensitive method to measure nuclear anapole moments~\cite{RahmlowPhD,Altuntas2018}. Efficient sources for the former isotopologues are highly sought after, as only odd isotopologues exhibit nuclear parity violation effects ~\cite{Kozlov1995}. An efficient buffer gas source like the one presented here could thus immediately facilitate measurements of parity violating matrix elements across several BaF isotopologues with Hz-level precision.

As for CaF~\cite{Truppe2017a}, a next step to achieve even higher molecular flux and lower initial temperatures would be to replace the $\mathrm{BaF}_2$ ablation with the ablation of a barium metal followed by a chemical reaction with $\textrm{SF}_6$ to form $\mathrm{BaF}$. This would also facilitate the use of isotope enriched barium sources to further boost the sensitivity in measurements of nuclear anapole moments. 

Moreover, the demonstrated optical cycling sets the stage for future transversal laser cooling of our BaF molecular beam. We can estimate that a total of $\sim 2000$ photons will have to be scattered for transversal laser cooling. Based on the vibrational branching this should be achievable with only two repumpers. Going one step further, laser slowing will immediately reveal the unknown additional branching ratio into the \deltastate state and hence the applicability of spontaneous forces for the laser cooling of BaF.

Looking beyond spontaneous forces, the high powers that can easily be generated at all relevant wavelengths, in conjunction with the observed narrow linewidth of the \deltastate state, open many interesting possibilities for the exploration of stimulated forces for slowing, cooling and trapping of molecules. 

\begin{acknowledgments}
We are indebted to Tilman Pfau and Jun Ye for generous support. We acknowledge valuable discussions with Yan Zhou, Timothy Steimle, Alejandra Collopy and Marc Scheffler. We thank Stefanie Barz and Martin Dressel for the loan of equipment. This project has received funding from the European Union's Horizon 2020 research and innovation programme under the Marie Sk\l odowska-Curie grant agreement No. 746525 (\textit{coolDips}), and from the Baden-W\"urttemberg Stiftung through the Eliteprogramme for Postdocs. The research of IQST is financially supported by the Ministry of Science, Research and Arts Baden-W\"urttemberg.
\end{acknowledgments}

\section*{Appendix}
\subsection{Franck-Condon factors}
We estimate the Franck-Condon factors (FCF) of the first and second excited states \exs and $B^2\Sigma$ based on the molecular constants from Ref.~\cite{Effantin1990}. The equilibrium distance $r_e$ is related to the rotational constant $B$ by $r_e = \sqrt{h/8\pi^2\mu B}$, where $\mu = 16.7\,$u is the reduced mass of BaF in atomic mass units. This yields $r_e = 2.1621\,$\AA, $r_e =2.1800\,$\AA\, $r_e =2.2040\,$\AA\, and $r_e =2.19382\,$\AA\,
for the $X$, $A$, $B$ and $A'\Delta$ states, respectively. Based on this we use Morse potentials and integrate the Schroedinger equation to find the molecular wavefunctions and their respective overlap integrals~\cite{Lopez2002,farkascomparat}. The results are shown in Table~\ref{tab:FCF}. While the $X-A$ FCFs for \baf138 are highly diagonal, the $X-B$ FCFs are not, due to the mismatch between the $X$ and $B$ equilibrium distances. 

\begin{table}[htb]
\begin{tabular*}{0.45\textwidth}{c @{\extracolsep{\fill}} cccc}

\multicolumn{5}{c}{$A^2\Pi_{1/2}(\nu')\rightarrow X^2\Sigma(\nu)$}\\
\hline 
\hline
 & $\nu'=0$ & $\nu'=1$ & $\nu'=2$ & $\nu'=3$ \\ 
\hline
$\nu=0$ & $0.9633$ & $0.0363$ & $4\times10^{-4}
$ & $5 \times 10^{-7}$\\ 
$\nu=1$ & $0.0356$ & $0.8900$ & $0.0730$ & $1.4 \times 10^{-3}$\\ 
$\nu=2$ & $1.1\times 10^{-3}$ & $0.0702$ & $0.8162$ & $0.1095$ \\ 
$\nu=3$ & $1\times10^{-5}$ & $3.4\times 10^{-3}$ & $0.1034$ & $0.7423$ \\ 
\end{tabular*} 
\vspace{15pt}

\begin{tabular*}{0.45\textwidth}{c @{\extracolsep{\fill}} cccc}
\multicolumn{5}{c}{$B^2\Sigma(\nu')\rightarrow X^2\Sigma(\nu)$}\\
\hline 
\hline
 & $\nu'=0$ & $\nu'=1$ & $\nu'=2$ & $\nu'=3$ \\ 
\hline 
$\nu=0$ & $0.8212$ & $0.1604$ & $0.0171$ & $1.3\times10^{-3}$ \\ 
$\nu=1$ & $0.1626$ & $0.5257$ & $0.2607$ & $0.0457$ \\ 
$\nu=2$ & $0.0153$ & $0.2689$ & $0.3104$ & $0.3123$ \\ 
$\nu=3$ & $9\times 10^{-4}$ & $0.0416$ & $0.3286$ & $0.1622$ \\  
\end{tabular*} 
\vspace{15pt}

\begin{tabular*}{0.45\textwidth}{c @{\extracolsep{\fill}} cccc}
\multicolumn{5}{c}{$A'\Delta_{3/2}(\nu')\rightarrow X^2\Sigma(\nu)$}\\
\hline 
\hline
 & $\nu'=0$ & $\nu'=1$ & $\nu'=2$ & $\nu'=3$ \\ 
\hline 
$\nu=0$ & $0.8721$ & $0.1178$ & $9.5\times 10^{-3}$ & $6\times10^{-4}$ \\ 
$\nu=1$ & $0.1206$ & $0.6462$ & $0.2044$ & $0.0264$ \\ 
$\nu=2$ & $7.1\times 10^{-3}$ & $0.2147$ & $0.4606$ & $0.2631$ \\ 
$\nu=3$ & $2\times 10^{-4}$ & $0.0204$ & $0.2843$ & $0.3121$ \\  
\end{tabular*} 

\caption{Franck-Condon factors of the $A^2\Pi_{1/2}(\nu')\rightarrow X^2\Sigma(\nu)$, $B^2\Sigma(\nu')\rightarrow X^2\Sigma(\nu)$ and $A'\Delta_{3/2}(\nu')\rightarrow X^2\Sigma(\nu)$ transitions.}
\label{tab:FCF}
\end{table}

\begin{table*}[ht]
\begin{tabular*}{0.95\textwidth}{l @{\extracolsep{\fill}} cccc}
& $ F=0 ,\:m_F=0$ & $ F=1 ,\:m_F=-1$ & $ F=1 ,\:m_F=0$ & $ F=1 ,\:m_F=1$ \\ \hline $
J=1/2,\:F=0 ,\:m_F=0 $ & $   0.0000$ & $    0.2222$ & $   0.2222$ & $   0.2222$ \\ $
J=1/2,\:F=1 ,\:m_F=-1$ & $   0.1282$ & $    0.2493$ & $   0.2493$ & $   0.0000$ \\ $
J=1/2,\:F=1 ,\:m_F=0 $ & $   0.1282$ & $    0.2493$ & $   0.0000$ & $   0.2493$ \\ $
J=1/2,\:F=1 ,\:m_F=1 $ & $   0.1282$ & $    0.0000$ & $   0.2493$ & $   0.2493$ \\ $
J=3/2,\:F=1 ,\:m_F=-1$ & $   0.2051$ & $    0.0007$ & $   0.0007$ & $   0.0000$ \\ $
J=3/2,\:F=1 ,\:m_F=0 $ & $   0.2051$ & $    0.0007$ & $   0.0000$ & $   0.0007$ \\ $
J=3/2,\:F=1 ,\:m_F=1 $ & $   0.2051$ & $    0.0000$ & $   0.0007$ & $   0.0007$ \\ $
J=3/2,\:F=2 ,\:m_F=-2$ & $   0.0000$ & $    0.1667$ & $   0.0000$ & $   0.0000$ \\ $
J=3/2,\:F=2 ,\:m_F=-1$ & $   0.0000$ & $    0.0833$ & $   0.0833$ & $   0.0000$ \\ $
J=3/2,\:F=2 ,\:m_F=0 $ & $   0.0000$ & $    0.0278$ & $   0.1111$ & $   0.0278$ \\ $
J=3/2,\:F=2 ,\:m_F=1 $ & $   0.0000$ & $    0.0000$ & $   0.0833$ & $   0.0833$ \\ $
J=3/2,\:F=2 ,\:m_F=2 $ & $   0.0000$ & $    0.0000$ & $   0.0000$ & $   0.1667$ \\ 
\end{tabular*}
\caption{Branching ratios for the $A^2\Pi_{1/2}$ positive parity state to the \gsshort, $N = 1$ state.}
\label{tab:branchingratios}
\end{table*}

\subsection{Energies and branching ratios}
As discussed in the main text, due to the coupling of the nuclear spin $I({}^{19}\mathrm{F})$ and the total angular momentum $J$ the \gs ground state of \baf138 splits up into four hyperfine levels. For a given rotational quantum number $N$ this leads to $F=N+1,N,N,N-1$. Solving the effective Hamiltonian following e.g. Ref.~\cite{Chen2016}, we find for the $N=1$ state, $F=2,1,0,1$ with energy shifts $+56.78\,$MHz, $+22.72\,$MHz, $-67.13\,$MHz and $-94.95\,$MHz relative to the center-of-mass energy. These levels can thus be addressed using equidistant sidebands generated by a single electro-optical modulator operating at $\sim 39\,$MHz. The hyperfine structure in the excited state contains two levels $F'=0,1$ that are not individually resolved.

The dipole transition branching ratios from the excited state $m_F'$ sublevels to the ground state $m_F$ sublevels are essential for the modeling of optical cycling. The calculation of these branching ratios follows the same strategy as described in Ref.~\cite{NorrgardPhD} for SrF, and requires both the excited and ground states to be represented in the same Hund's case basis. It is important to note that the ground state hyperfine levels with $F=N$ are superpositions of the two possible $J$ values for a given rotational level $N$. It is common to denote these states by $F=N^+$ and $F=N^-$. For the  $N=1$ state these superpositions are given by
\begin{align}
| F=1^-\rangle &= -\beta_1|J=\frac{3}{2},F=1\rangle+\alpha_1 |J=\frac{1}{2}, F=1\rangle\nonumber\\
| F=1^+\rangle &= \alpha_1|J=\frac{3}{2},F=1\rangle+\beta_1 |J=\frac{1}{2}, F=1\rangle
\label{eq:superpositions}
\end{align}
with the coefficients $\alpha_1=0.9593$, $\beta_1=0.2824$, $\alpha_2=0.9858$ and $\beta_2 = 0.1679$. Using this and following Refs.~\cite{NorrgardPhD,Chen2016} we summarize our branching ratios in Table~\ref{tab:branchingratios}. Note the difference of our results from the branching ratios reported in Ref.~\cite{Chen2016}, which stems from an incorrect choice of sign in the equivalent of Eq.~\ref{eq:superpositions} in that work. However, the difference does not significantly alter any of the results of the rate equation model discussed in the main text~\cite{Chen2017}. 

\newpage
\bibliography{biblio}
\end{document}